\documentclass[twocolumn,twocolappendix,trackchanges]{aastex631}

\usepackage{booktabs}
\usepackage{graphicx}	
\usepackage{amsmath}

\newcommand{\SatGen}{{\textsc{SatGen}}}
\newcommand{\streamdf}{{\textsc{streamdf}}}
\newcommand{\Galpy}{{\textsc{Galpy}}}
\newcommand{\Gala}{{\textsc{Gala}}}

\newcommand{\apjbf}[1]{#1}

\begin{document}
\reportnum{YITP-SB-2024-32}

\title{Semi-Analytic Modeling of Dark Matter Subhalo Encounters with Thin Stellar Streams: \\ 
Statistical Predictions for GD-1-like Streams in CDM}

\author{Duncan K. Adams}
\affiliation{C.N.~Yang Institute for Theoretical Physics, Stony Brook University, NY 11794, USA}

\author{Aditya Parikh}
\affiliation{C.N.~Yang Institute for Theoretical Physics, Stony Brook University, NY 11794, USA}

\author{Oren Slone}
\affiliation{C.N.~Yang Institute for Theoretical Physics, Stony Brook University, NY 11794, USA}

\author{Rouven Essig}
\affiliation{C.N.~Yang Institute for Theoretical Physics, Stony Brook University, NY 11794, USA}

\author{Manoj Kaplinghat}
\affiliation{Department of Physics and Astronomy, University of California, Irvine, CA 92697 USA}

\author{Adrian M. Price-Whelan}
\affiliation{Center for Computational Astrophysics, Flatiron Institute, 162 Fifth Ave., New York, 10010, NY, USA}

\begin{abstract}

Stellar streams from disrupted globular clusters are dynamically cold structures that are sensitive to perturbations from dark matter subhalos, allowing them in principle to trace the dark matter substructure in the Milky Way.
We model, within the context of $\Lambda$CDM, the likelihood of dark matter subhalos to produce a significant feature in a GD-1-like stream and analyze the properties of such subhalos. 
We generate \apjbf{many} realizations of the subhalo population within a Milky Way mass host halo \apjbf{using the semi-analytic code \SatGen{}}, accounting for \apjbf{effects such as} tidal stripping and dynamical friction. 
The subhalo distributions are combined with a GD-1-like stream model, and the impact of subhalos that pass close to the stream are modeled with \Gala{}. 
We find that subhalos with masses in the range \apjbf{$2\times 10^6 M_{\odot} - 10^8 M_{\odot}$} at the time of the stream--subhalo encounter, corresponding to masses of about \apjbf{$2 \times 10^7 M_{\odot} - 10^9 M_{\odot}$} at the time of infall, are the likeliest to produce gaps in a GD-1-like stream. 
We find that gaps occur on average \apjbf{$\sim$3}~times per 
realization of the host system.  These gaps have typical widths of \apjbf{$\sim(5 - 27)$}~deg and fractional underdensities of $\sim (10 - 30)\%$, with larger gaps being caused by \apjbf{heavier} subhalos.  
The stream--subhalo encounters responsible for these have impact parameters $(0.1 - 1.5)$~kpc and relative velocities \apjbf{$\sim(200 - 410)$}~km/s. 
\apjbf{We also investigate the effects of increasing the host-halo mass on the gap properties and formation rate.}
\end{abstract}


\section{Introduction} 
\label{sec:intro}

In the $\Lambda$-Cold-Dark-Matter ($\Lambda$CDM) cosmological paradigm, structure formation in the early Universe proceeds hierarchically, implying the existence of bound DM substructures 
well below the threshold of galaxy formation
\citep[e.g.,][]{Moore:1998}. 
Alternative DM paradigms often instead predict different subhalo mass functions, with no expected substructure (or at least significantly depleted number density) below some mass scale.
For example, warm dark matter (WDM) erases power on small scales in the early universe \citep[e.g.,][]{Bode:2001} and self- or other interacting dark matter (SIDM) may experience similar free-streaming effects along with late-time evolution that changes the internal structure of SIDM subhalos relative to CDM counterparts \citep[e.g.,][]{Spergel:2000, Adhikari:2022sbh, Roy:2023}.
Identifying and characterizing the existence and population properties of low-mass DM subhalos would therefore provide an important test of DM models. However, low mass subhalos (i.e., $M\lesssim 10^8$--$10^9~\textrm{M}_\odot$ in peak halo mass) are not expected to contain luminous matter because of a variety of effects that impact their ability to retain gas and form stars (predominantly reionization, gas cooling, and tidal stripping; e.g., \citealt{Quinn:1996, Bullock:2000wn, Okamoto:2008, Finlator:2017, Grand:2021}). 
To detect and study DM subhalos below this threshold for galaxy formation, one therefore has to rely on indirect methods.
Two promising paths towards astrophysical characterization of low-mass DM subhalos are  gravitational lensing and gravitational interactions with stars and gas.
In the context of lensing, DM subhalos may be detectable through perturbations to strong lensing images of background galaxies \citep{Vegetti:2014, Minor:2020bmp, Minor:2020hic, Vegetti:2023, Zhang:2023wda}, 
flux ratio anomalies in lensed quasar systems \citep{Dalal:2001fq,Gilman:2024},  
through weak lensing effects on background stars \citep{KVT:2018, Mondino:2020, Mondino:2024}, 
or in future high-resolution cosmic-microwave-background lensing measurements~\citep{Nguyen:2017zqu,CMB-HD:2022bsz,MacInnis:2024znd}.

In the context of gravitational interactions with stars and gas, various ideas have been proposed such as as the heating of the galactic disk of stars or gas~\citep{Carr:1997cn, Velazquez:1998pz, Ardi:2002ee, Hayashi:2006bg, Kazantzidis:2007hy, Hopkins:2008vc, Kazantzidis:2009zq}, heating of stars in ultrafaint dwarfs~\citep{Dalal:2022rmp, Graham:2024hah}, survival of weakly-bound stellar binaries~\citep{Penarrubia:2010pa} and perturbations of stellar streams. 

The typical velocity impulse imparted by a single passing subhalo is small. Using the impulse approximation and assuming subhalos with a Navarro-Frenk-White (NFW) \citep{Navarro:1996} mass profile, the expected velocity signal is
\begin{equation}
    \Delta v \sim 1~\textrm{km}~\textrm{s}^{-1} \, 
        \left(\frac{M_{\textrm{sub}}}{10^7~\textrm{M}_\odot}\right)^{2/3} \, 
        \left(\frac{v_{\rm rel}}{100~\textrm{km}~\textrm{s}^{-1}}\right)^{-1}
\end{equation}
for a direct impact of a subhalo of mass $M_{\textrm{sub}}$ and relative velocity $v_{\rm rel}$, assuming a typical impact parameter equal to the scale radius of the subhalo \citep[see, e.g.,][]{Erkal:2015b, Bonaca:2024}.
This means that subhalos are only expected to be detectable in dynamically cold systems like wide binary stars \citep{Yoo:2004, Brandt:2016, Penarrubia:2019} and stellar streams \citep{Ibata:2002, Johnston:2002}.
In this article, we focus on the impact of DM subhalos on stellar streams.

Stellar streams form from tidally disrupted globular clusters or dwarf galaxies as they orbit within the Milky Way's (MW's) gravitational potential \citep[see, e.g.,][]{Bonaca:2024}. 
Streams originating from globular clusters (``thin'' stellar streams) have small velocity dispersions and widths, making them sensitive to perturbations from DM subhalos \citep{Johnston:2002, Ibata:2002, Yoon:2011, 2017MNRAS.466..628B, 2024ApJ...975..135C}.
For example, models of thin streams predict that typical velocity dispersions can be in the range $\sim (0.1$--$1)~\textrm{km}~\textrm{s}^{-1}$, comparable to the velocity dispersions of the outskirts of globular clusters \citep{Baumgardt:2018} and to the expected velocity perturbations from low-mass subhalos.
Interactions between streams and subhalos can produce characteristic features such as density variations or ``gaps'' (under-densities) along the stream \citep{Carlberg:2012, Erkal:2015a, Erkal:2016, Sanders:2016}, as well as ``off-track'' features such as spurs \citep{Yoon:2011, Bonaca:2019}, depending on the properties of the perturber and the geometry of the interaction.
Weak encounters with subhalos (i.e., larger relative velocity, lower mass, or larger impact parameter) generally produce gaps, whereas strong encounters (i.e., smaller relative velocity, higher mass, or smaller impact parameter) can produce more dramatic features that split a stream or spray stars out of the main track of the stream.

Although photometric surveys yielded the first clues of non-trivial density structures in streams \citep{2003AJ....126.2385O,2006ApJ...643L..17G}, astrometric data from the  Gaia Space Telescope revealed clear gaps and off-track features in GD-1 \citep{2018ApJ...863L..20P}. Furthermore, follow up spectroscopy on specific streams (e.g. Aliqa-Uma) \citep{2019ApJ...881L..37B,2021ApJ...911..149L} has revealed a diverse landscape of morphological features in stellar streams, motivating a need to better understand the rates of stream--subhalo encounters and their effects on stream morphology.

In particular, connecting the underlying subhalo population of the host galaxy (e.g., the MW) to concrete predictions for gap formation rates and expected gap properties has not been fully solved. Analytic estimates \citep{Yoon:2011} suggest that $\mathcal{O}(10)$ close encounters with subhalos with masses between $10^6 M_\odot - 10^9 M_\odot$ are expected for streams such as GD-1. However, not all close encounters lead to significant gaps, and by combining models for gap growth based on circular orbits with analytic subhalo distributions, \citep{Erkal:2016} estimated an $\mathcal{O}(1)$ number of gaps in GD-1. To refine these estimates, models of streams that account for eccentric orbits must be combined with more realistic subhalo populations. In particular, semi-analytic models are well suited to this task, since they are capable of rapidly producing many realizations of subhalo populations, while zoom-in simulations currently do not reach resolutions required to reliably predict subhalo populations for masses below $\sim 10^6 M_\odot$. Semi-analytic subhalo populations were applied to study Pal-5 in \citep{2024arXiv240611989M}, however that study \apjbf{assumed a circular orbit for Pal-5} and did not perform detailed simulations of stream--subhalo encounters.

\begin{figure*}[t]
    \centering
\includegraphics[width=0.8\textwidth]{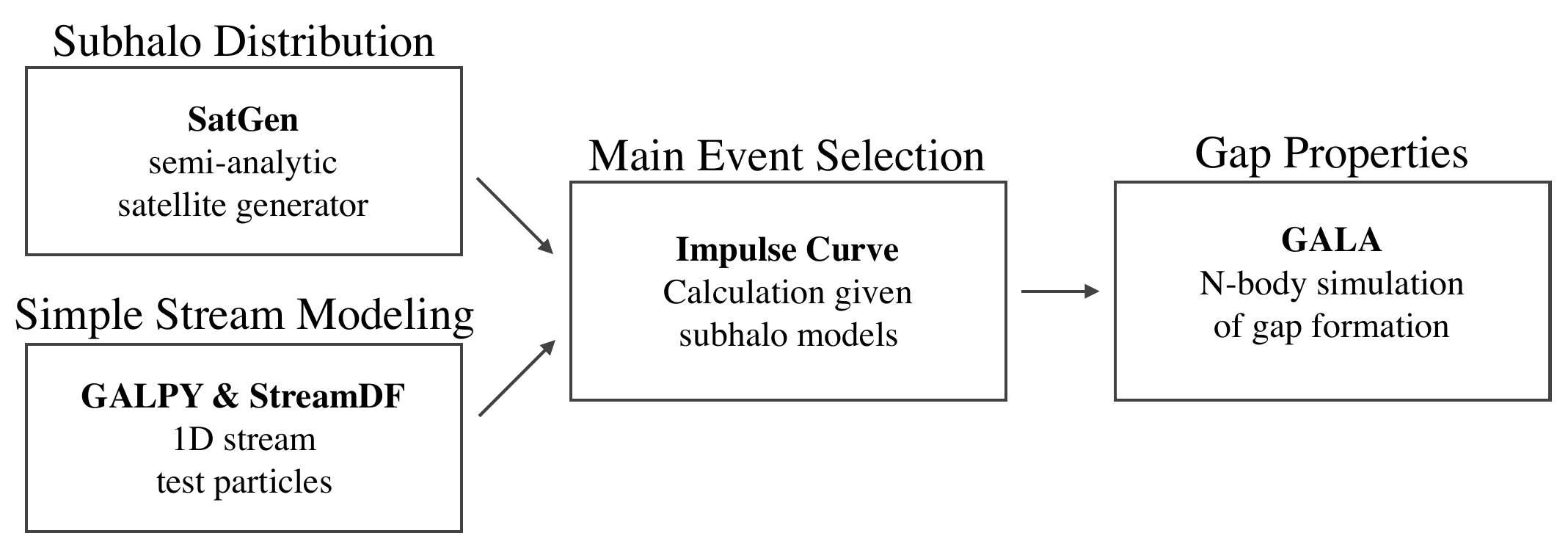}
    \caption{A schematic representation of our implementation and analysis workflow emphasizing the modularity of our approach in this study. We first use \SatGen{} to generate subhalo populations for MW mass galaxies. Simultaneously, we use the \streamdf{} module in \Galpy{} to generate the orbital history of a GD-1-like stream. Taking these outputs, we then compute impulse curves and use them as a preliminary criteria to identify potential stream--subhalo encounters of interest, i.e. those that may produce sizable gaps in a GD-1-like stream. Finally, we simulate these events in \Gala{} and isolate gap properties such as width and depth, as well as encounter properties such as the subhalo's instantaneous and infall masses, the impact parameter and relative velocity, and the time of impact. Each of these modules is largely self-contained, and as a result, this analysis workflow is easily amenable to changes within each module, e.g., using a different stream model. The steps of this analysis workflow are described in further detail in Section~\ref{sec:methodology}.}
    \label{fig:schematic}
\end{figure*}

Motivated by these considerations, as well as the wealth of diverse stream morphologies observed in the MW, the aim of this work is to develop a robust statistical understanding of the subhalo properties that allow for gap formation in a GD-1-like stellar stream, as well as the frequency of such gaps.
We combine semi-analytic modeling of MW subhalo populations together with simulations of a GD-1-like stellar stream, to quantify distributions of the salient properties of stream--subhalo encounters that can form significant gaps. Our method can easily be extended to other streams or non-MW host galaxies. Additionally, while this study considers the case of collisionless CDM, the approach can be adapted to other DM paradigms including those with self-interactions or dissipative processes, \apjbf{provided that the relevant physics for subhalo evolution in these paradigms is accounted for}.

The paper is organized as follows. Section~\ref{sec:methodology} discusses the methods employed and our analysis, including the semi-analytical treatment of stream--subhalo encounters as well as the detailed analysis of gap formation. Section~\ref{sec:results} presents our results, focusing on the properties of encounters that form significant gaps as well as their expected frequencies. Section~\ref{sec:discussion} discusses our main results, compares it to other work in the literature,  mentions some of its limitations and future improvements, and also discusses future applications. Section~\ref{sec:conclusions} concludes this paper.
Three appendices provide additional information. 

\section{Methodology} 
\label{sec:methodology}

The goal of this work is to provide a robust framework to estimate statistical properties of stream features while accounting for uncertainties in modeling of the host galaxy (in our case the MW) and the properties of the stream--progenitor system. The approach presented here is to initially sample a probability distribution of relevant MW subhalo properties, such as their masses and concentrations, together with the locations and velocities of the subhalos and stream stars at a given time, and then to identify which stream--subhalo encounters are capable of producing a sizable perturbation to the stream's density profile. While this study focuses predominantly on gap formation, the formalism could straightforwardly be extended to additional features such as the formation of spurs or more subtle effects that would appear in the stream density power spectrum.

We sample the initial properties of subhalos using the \SatGen{} ~\citep{Jiang:2020rdj,2022MNRAS.509.2624G} semi-analytical satellite generator, which is able to rapidly create large ensembles of MW mass systems and vary modeling assumptions. 
This is useful to understand the systematics that can affect statistical properties of stream features, which is a goal of this study. 
For each sampled subhalo, we calculate the impulse curve along a mock stream that has also been evolved in a simplified MW potential and project this impulse curve onto the direction parallel to the stream's orbit. This ``parallel impulse curve'' allows one to identify the most likely events that could cause sizable gaps in the stream, and those events are then simulated using the \Gala{} code~\citep{GALA} in order to identify detailed properties of the gap formation. Events that are deemed interesting are saved and collected for statistical analysis. A schematic of the pipeline is shown in Figure~\ref{fig:schematic} and 
presented in further detail in the subsections below.

\subsection{Subhalo Sampling from a Semi-Analytic Satellite Generator}
\label{sec:semianalytic} 

The \SatGen{} software package~\citep{Jiang:2020rdj,2022MNRAS.509.2624G} is used to create realizations of the CDM subhalo population around a MW mass host. The package combines semi-analytic prescriptions for galaxy evolution encompassing merger trees, subhalo structural evolution, and dynamical effects such as tidal stripping, tidal tracks, and dynamical friction. 
The \SatGen{} code populates the initial subhalo catalog via a merger tree as detailed in \citep{2008MNRAS.383..557P}. Given the host halo's final mass (and the required mass resolution), the algorithm constructs a mass assembly history in a series of redshift intervals using the extended Press-Schecter formalism \citep{1991ApJ...379..440B,1991MNRAS.248..332B} to determine when subhalos above the mass resolution merge into the host. It then recursively repeats this procedure to construct a mass assembly history for each subhalo on the same set of redshift intervals.

The output is a merger tree with subhalo infall masses and redshifts. For each subhalo, an infall concentration is determined based on its accretion history following the prescription in~\citep{2009ApJ...707..354Z}, 
and orbital parameters are assigned using the distributions from~\citep{Li:2020bom}. 
The orbital and mass loss history of each subhalo is captured via an orbit integration scheme wherein each subhalo is evolved in the potential of its parent, but can also become unbound as a result of tidal forces from the grandparent. The orbit integration accounts for dynamical friction, tidal stripping, and tidal tracks. Subhalos undergoing mass loss are modeled using the Green profile~\citep{Green:2019zkz} whose parameters are the instantaneous (time-dependent) subhalo mass $M_{\rm sub}$, its concentration $c_{\rm sub}$, and a truncation radius $r_t$ that sets the mass truncation from the profile's outer region.

The host potential (in our case a MW mass system) is modeled as the sum of an NFW DM halo with virial mass $M_h$ and concentration $c_h$ determined from the assembly history using the results of ~\citep{2009ApJ...707..354Z},  together with a Miyamoto-Nagai stellar disk~\citep{1975PASJ...27..533M},
\begin{equation}
    \Phi^{\textrm{MN}}(R, z, t) = -\frac{GM_d(t)}{\sqrt{R^2 + \bigg(\sqrt{z^2 + b_d(t)^2} + a_d(t)\bigg)^2}} \,,
\end{equation}
with time-dependent mass and scale factors that are power-laws with respect to $M_h(t)$, see~\citep{2022MNRAS.509.2624G},
\begin{equation}
    \frac{X_d(t)}{X_d(t_0)} = \bigg[\frac{M_h(t)}{M_h(t_0)} \bigg]^{\beta_X}\, ,
\end{equation} 
where $X\equiv\,\{M,\,a,\,\text{or}\,\,b\}$, and $t_0$ corresponds to redshift zero. The values of the parameters used in this study are given in Table~\ref{tab:satgen_parameters}. Appendix \ref{sec:rot-curve} presents rotation curves for the \SatGen{} hosts used in this study and a discussion regarding how the disk mass was chosen.

This work uses a ``fiducial'' set of 400 MW mass halos created by \SatGen{} with the model parameters specified in Table~\ref{tab:satgen_parameters}. 
An additional ``high-mass'' sample of \apjbf{200}~halos, with the same set of parameters as the fiducial sample except for a MW halo mass that is twice as large, is used as a comparison point to bracket uncertainties in MW halo modeling and to study systematic effects on the results.

\begin{table}[t]
    \centering
    \begin{tabular}{l l l}
        \toprule
        Parameter & Value & Description \\
        \midrule
        $M_{\mathrm{res}}$ & $10^{5.8} M_\odot$ & Merger tree mass resolution\\
        $M_h(t_0)$ & $1~(2)\times10^{12} M_\odot$ & Host halo virial mass at $t_0$ \\ 
        $M_d(t_0)$ & $8.6 \times 10^{10} M_\odot$ & Disk mass at $t_0$ \\
        $a_d(t_0)$ & \apjbf{3.24~(4.07)~kpc} & Disk horizontal scale at $t_0$ \\
        $b_d(t_0)$ & \apjbf{0.26~(0.33)~kpc} & Disk vertical scale at $t_0$ \\
        $\beta_M$ & 1.0 & Disk mass growth index \\
        $\beta_a$ & 1/3 & Horizontal scale growth index\\
        $\beta_b$ & 1/3 & Vertical scale growth index\\
        \bottomrule
        
    \end{tabular}
    \caption{Parameters used in our sample of \SatGen{} Milky Way realizations. The mass of the Milky Way halo is chosen to be representative of the halo masses reported in Table 8 of \citep{2016ARA&A..54..529B}. We also create a sample of high-mass Milky Way halos whose \apjbf{parameters at $t_0$ are given in parentheses.}}
    \label{tab:satgen_parameters}
\end{table}

\begin{figure*}[t]
    \centering
    \includegraphics[width=2\columnwidth]{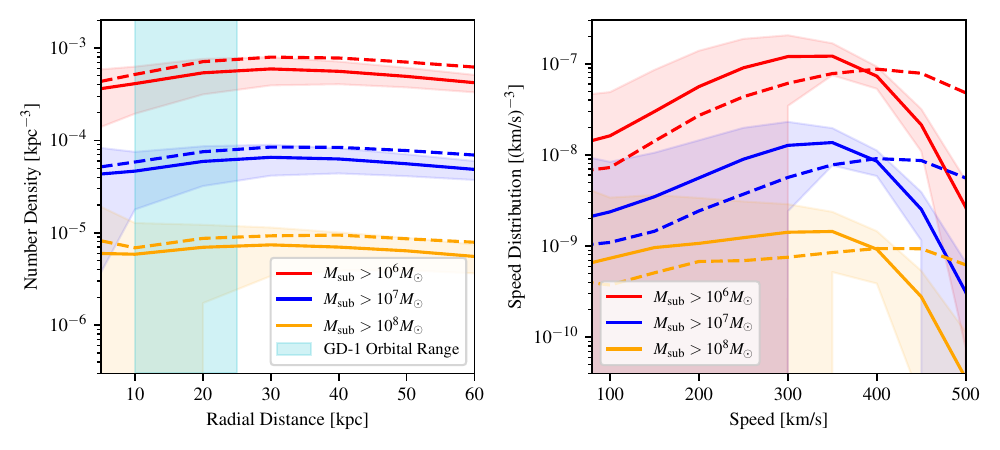}   
    \caption{\textbf{Left:} Subhalo number density versus distance from the center of a MW mass host galaxy, averaged over realizations. Solid curves and shaded bands represent the means and standard deviations of the fiducial sample of halos, respectively; dashed curves represent the means of the high-mass sample. The shaded vertical region indicates the approximate range where the GD-1 stream orbits within the MW. These curves are computed by counting the number of subhalos within spherical shells of width 5~kpc centered at a radial distance from the Galactic Center and dividing by the corresponding shell volume, mirroring the definition in~\citep{2023MNRAS.523..428B}.  \textbf{Right:} 
    The distribution  of the speeds of the subhalos in the galactic rest frame, with the same notations as the left panel. Curves are computed by counting the number of subhalos within spherical shells of width 100~km/s.}
    \label{fig:subhalo-density}
\end{figure*}

The left panel of Figure~\ref{fig:subhalo-density} shows subhalo number densities versus distance from the Galactic Center for both the fiducial and high-mass runs averaged from redshift~$0$ to redshift~$0.15$. Each color in the figure corresponds to a different range of subhalo masses. Notably, in the region where a GD-1-like stream resides throughout its lifetime, the subhalo number densities are weak functions of distance, and the difference between the fiducial and high-mass runs is of order $\sim30\%$. Within 50 kpc of the Galactic Center, these densities on average yield approximately 250 (350) subhalos between $(10^6-10^7) M_\odot$,  30 (40) subhalos between $(10^7-10^8) M_\odot$, and 4 (5) subhalos greater than or equal to $10^8 M_\odot$, in the fiducial (high-mass) samples. The right panel of the figure shows the speed 
distributions of subhalos in the same subhalo mass bins 
for both the fiducial and high-mass runs. The main effect of increasing the mass of the MW is additional support at high speeds, as would be expected from a MW halo with increased virial velocity, since $V_{\rm virial}\propto M_{\rm virial}^{1/3}$. Results of the high-mass run are discussed further below and also shown in Appendix~\ref{sec:app-highmass}.

Figure~\ref{fig:subhalo-density} highlights the simplicity of the final result, 
despite the highly non-trivial inner workings of the \SatGen{} code. The procedure developed in this work requires the multidimensional distribution function of subhalo properties (such as masses, density profile parameters, positions, and velocities) in a small region of phase space where the stream resides. Using this language is useful for developing intuition. For example, it is straightforward to estimate the expected change in gap properties in a GD-1-like stream due to an increase in the host halo mass. 
A larger halo mass corresponds to a slight increase in the number of subhalos in the region where GD-1 resides, 
which implies more potential perturbers and an increase in the number of expected gaps. Additionally, these perturbers have, on average, larger typical velocities (due to the larger virial velocity), leading to an additional increase in the encounter rate, which is linear in the velocity dispersion of 
perturbing subhalos~\citep{Yoon:2011}. Both effects can easily be understood in terms of the average encounter rate being proportional to the number density times velocity of perturbers. In principle, the larger velocity dispersion can also reduce the expected size of gaps, since the impulse onto the stream stars is inversely proportional to subhalo velocity.  However, we find that the expected gap properties are not significantly altered (see Appendix~\ref{sec:app-highmass}), consistent with what was shown in~\citep{2017MNRAS.466..628B}. 

\subsection{Stream--Subhalo Encounters and Event Selection}

Given the probability distribution of subhalos from the \SatGen{} code, we simulate a mock stream around the MW that is synchronized with \SatGen{}'s cosmic time. In this study, the mock stream is generated to mimic a GD-1-like stream using the \streamdf{} module from the \Galpy{} package \citep{2015ApJS..216...29B}. Given a host potential, a globular cluster progenitor orbit, the progenitor velocity dispersion, and the time elapsed since disruption began, \streamdf{} computes and outputs the positions and velocities of 2002 test particles along the stream track, as a function of time. Note that these test particles do not simply evolve along their orbit between timesteps, since the stream changes length throughout its evolution. \apjbf{Instead, the particles should be thought of as sampling the \emph{phase-space configuration} 
at various points along the stream, which are independently resampled for each time step.}

The GD-1-like stream in this study is modeled following the work of 
~\citep{2014ApJ...795...95B}. The progenitor is a globular cluster with velocity dispersion $\sigma_v = 0.365$~km/s, which began tidally disrupting 5.011~Gyr ago, with orbital parameters at redshift zero given in cylindrical coordinates by
\begin{eqnarray}
\{R,R\phi,z\} & = & \{12.492,1.501,7.098\}\text{ kpc} \\ 
\{v_R,R v_\phi,v_z\} & = & \{77.179,-254.059,-104.969\}\text{ km/s}. \nonumber
\end{eqnarray}

\apjbf{The stream is evolved separately in each host realization and the 3D locations of the stream test particles are obtained for each \SatGen{} snapshot, which are approximately 60 Myrs apart.} Then, a cubic spline function is fit \apjbf{for each particle's position and velocity across all snapshots}, and the resulting interpolation function is used to evaluate test particle locations on a much finer time grid.\footnote{\apjbf{The spline interpolation leads to a percent level error of the position and velocity of the stream track.}} To obtain the subhalo locations on the same time grid, we take two consecutive \SatGen{} snapshots and compute the subhalo orbital trajectory between them by \apjbf{solving the equations of motion in the host potential, assuming that the potential does not change during the 60~Myrs, which is an excellent approximation.} 
This procedure, performed in the range (0.5--4.5)~Gyrs ago, provides subhalo and stream locations and velocities over a finely sampled grid in time, enabling accurate tracking of close encounters.

Crude initial cuts are then placed to identify the most interesting encounters for further analysis.
A subhalo is selected for further analysis if it approaches within a  distance of five times its initial (infall) scale radius. This cut is particularly useful in removing unwanted encounters, since it is dynamically scaled for each subhalo given that more massive subhalos, with correspondingly larger scale radii, can more easily impart larger velocity kicks on stream particles. For each close approach we determine the duration of the encounter in units of \SatGen{} snapshots by linearly interpolating the subhalo's position between snapshots to find the time at which it enters and exits the distance threshold specified above. We then define two times, $t_\mathrm{pre}$ at the start of the entry snapshot and $t_\mathrm{post}$ at the end of the exit snapshot. We require \apjbf{$(t_\mathrm{post}-t_\mathrm{pre}) \leq 250$ Myr to remove encounters that only change the orbit adiabatically.}

\begin{figure}[t]
    \centering
    \includegraphics[width=0.75\columnwidth]{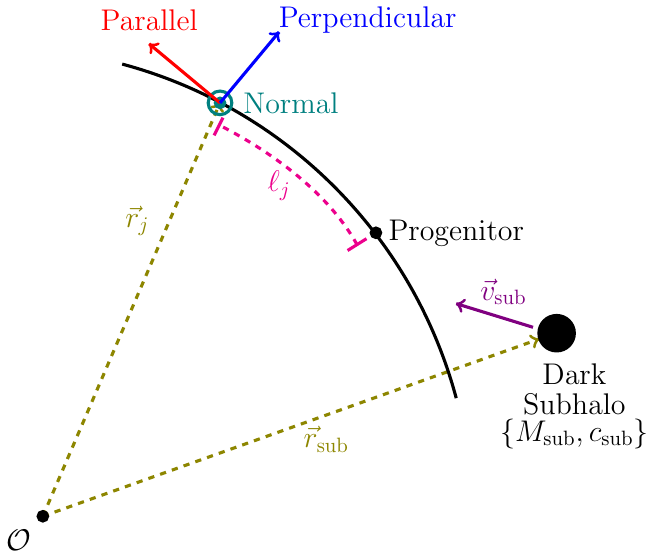}
    \caption{A schematic diagram of a stream--subhalo encounter and notations used in this study. Each dark subhalo is modeled by a Green profile~\citep{Green:2019zkz} with instantaneous mass $M_{\rm sub}$ and concentration $c_{\rm sub}$ (each subhalo also has a mass loss fraction parameter that sets mass truncation from the outskirts of subhalo, not shown here). Subhalo locations and velocities are denoted by $r_{\rm sub}$ and $\vec{v}_{\rm sub}$ relative to the Galactic Center (denoted $\mathcal{O}$). Locations along the stream (for each test particle $j$) are given by vectors $\vec{r}_{j}$, and the arclength between the progenitor and location $j$ is denoted by $\ell_j$. The red arrow shows the direction locally parallel to stream particle $j$, used to project the local impulse from the encounter in the direction parallel to the stream. Also shown are perpendicular and normal directions, which can be used to analyze other stream features, such as spurs. The black curve is a schematic of the stream track. Note that the leading and trailing arms are not necessarily equal in length due to orbital modulation of the stream length.
    }
    \label{fig:stream_schematic}
\end{figure}

The next step in the analysis is to identify stream--subhalo encounters that have higher probabilities of creating stream gaps. This is done by calculating impulse curves along the stream and projecting them onto the direction parallel to the stream~\citep{Erkal:2015a} (at this point in the analysis, the stream is modeled as a 1D curve, and therefore this procedure is well defined). A schematic diagram of a stream--subhalo interaction and the corresponding coordinate system we defined is shown in Figure~\ref{fig:stream_schematic}. Other features, such as spur formation, could be studied by also considering impulses in the perpendicular directions; however, this is beyond the scope of the current work and will be considered in follow-up studies. The shape of this parallel impulse curve provides additional information that can be used to select the most interesting events.

\begin{figure}[t]
    \centering
    \includegraphics[width=\columnwidth]{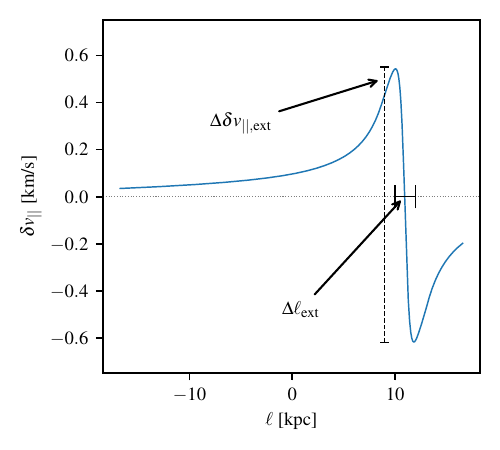}
    \caption{Representative example of $\delta v_{||}$ as a function of displacement along the stream from the progenitor location caused by a close stream--subhalo encounter. As discussed in the main text and in considerable detail in \citep{Erkal:2015a}, the shape of this curve determines the rate of gap growth and the location of the gap along the stream. Also shown are the peak-to-trough difference in velocity $\Delta \delta v_{{||,{\rm ext}}}$, and the arclength between the peak and trough $\Delta \ell_{\rm ext}$.}
    \label{fig:delta-v}
\end{figure}

For each stream--subhalo encounter that passes the initial cuts, the impulse curve is calculated for each stream test particle $j$, by integrating the gravitational acceleration over time as follows,
\begin{equation}
\delta \vec{v}_j = \int_{t_\mathrm{pre}}^{t_\mathrm{post}} G \frac{M_{\rm sub,enc}(r)}{r^2} \hat{r} dt \ ,
\label{eq:impulse}
\end{equation}
where $G$ is Newton's constant, $M_{\rm sub,enc}(r)$ is the subhalo's enclosed instantaneous mass within radius $r$, and $\vec{r} \equiv \vec{r}_{j} - \vec{r}_{\rm sub}$ is the vector pointing between the subhalo's center and stream test particle $j$'s location. 
The $\delta \vec{v}_j$ values are then projected onto the direction parallel to the stream at location $\vec{r}_j$, and the $\vec{r}_j$ location is associated with an arclength, $\ell_j$, along the stream track (from the progenitor to the test particle $j$ at the time of closest approach). The final result is a curve of the form $\delta v_{||}(\ell)$. Note that Equation~\eqref{eq:impulse} is frame independent. Also note that an assumption at this stage in the analysis is that subhalo acceleration does not alter the stream's trajectory throughout the encounter; this assumption is relaxed in Section~\ref{sec:Gala_sim}. The final result of the procedure is a curve such as that shown in Figure~\ref{fig:delta-v}, which can be used to estimate gap-formation properties.

Intuitively, a stream--subhalo encounter forms a gap because stars on either side of the point of closest approach get pulled towards each other. The prograde (retrograde) acceleration increases (decreases) the orbital period of affected stars, leading to a secular drifting apart of stars near the impact site. Figure~\ref{fig:delta-v} shows an example $\delta v_{||}(\ell)$ curve for a candidate stream--subhalo encounter. Such curves were studied in detail in~\citep{Erkal:2015a} for idealized systems, and much of the intuition from that paper carries over to this study. The characteristic shape of $\delta v_{||}(\ell)$ curves that lead to significant gap formation have two extrema at $\ell_{{\rm ext},1}$ and $\ell_{{\rm ext},2}$ with relatively small separation in arclength $\Delta \ell_{\rm ext} \equiv |\ell_{{\rm ext},2}-\ell_{{\rm ext},1}|$, and large absolute values of impulse difference $\Delta \delta v_{{||,{\rm ext}}} \equiv |\delta\vec{v}_{||}(\ell_{{\rm ext},2}) - \delta\vec{v}_{||}(\ell_{{\rm ext},1})|$. $\Delta \ell_{\rm ext}$ sets the initial size of the gap region, while $\Delta \delta v_{{||,{\rm ext}}}$ sets the rate of secular gap growth (the gap size varies as the stream orbits due to interstellar distances between stream members compressing and stretching throughout the orbit). 
 We place additional cuts on stream--subhalo encounters, requiring the $\delta v_{||}(\ell)$ curve to exhibit exactly two extrema, with $\Delta\ell_{\rm ext}/\ell_{\rm stream} \leq 0.4$ (where $\ell_{\rm stream}$ is the total arclength of the stream at the time of closest approach). \apjbf{These requirements select for events that have $\delta v_{||}(\ell)$ curves with the correct structure to lead to gap formation}. We also require $\Delta \delta v_{{||,{\rm ext}}} \geq 0.1$ km/s \apjbf{, ensuring that the perturbation is large enough to overcome the velocity dispersion of the stream.} 
Encounters satisfying these additional cuts are then analyzed in further detail using dedicated N-body stream simulations.

\subsection{Final Event Selection and Gap Properties}
\label{sec:Gala_sim}

\begin{figure*}
    \centering
    \includegraphics[width=2\columnwidth]{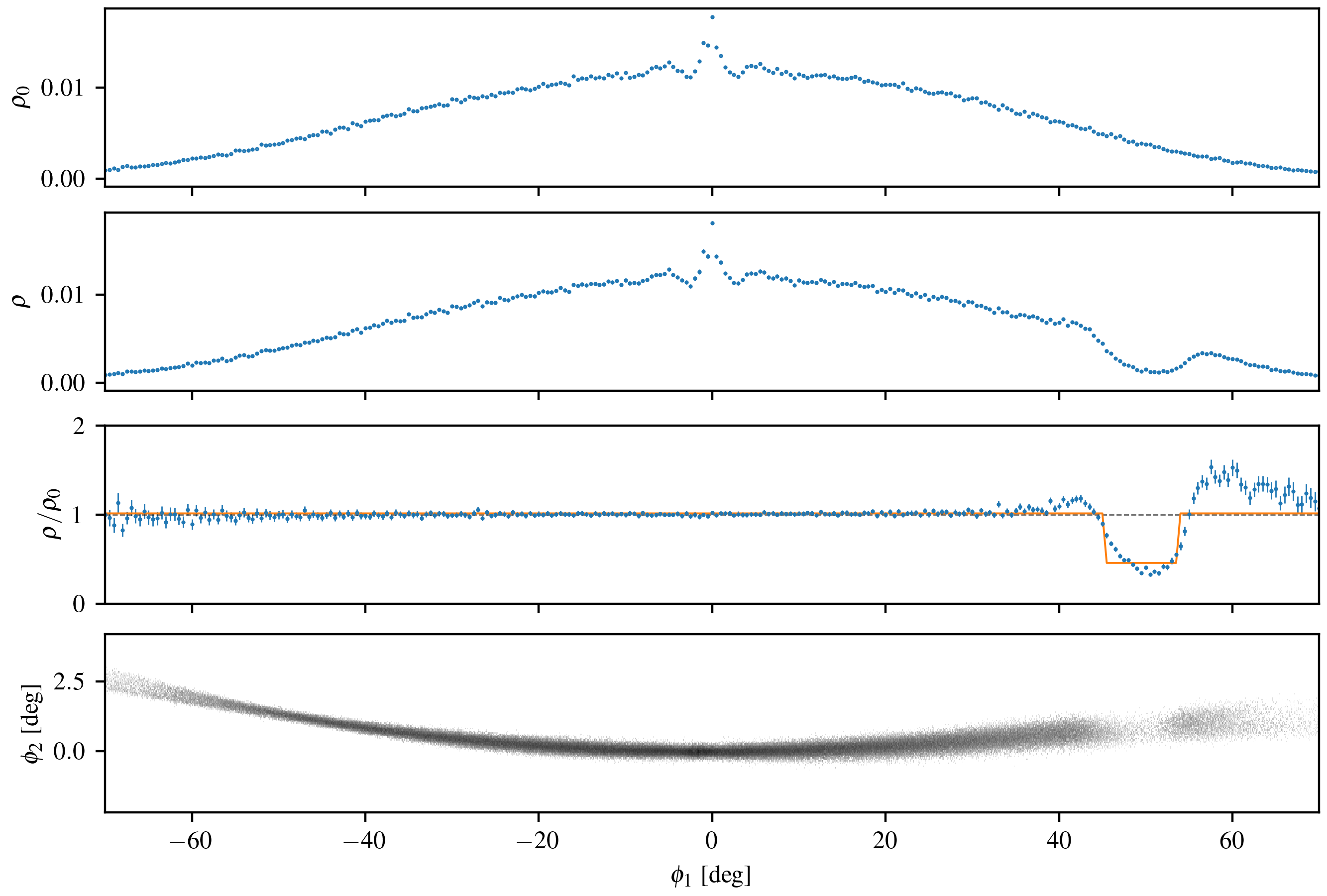}
    \caption{Output of the \Gala{} stream simulation for the candidate event whose $\delta v_{||}$ curve is shown in Figure~\ref{fig:delta-v}. All results are shown as functions of $\phi_1$, the angular coordinate along the stream trajectory, and binned into $0.1$ deg intervals in $\phi_1$. \textbf{First Panel:} Density of the hypothetical, unperturbed stream, $\rho_0(\phi_1)$. \textbf{Second Panel:} Density of the perturbed stream, $\rho(\phi_1)$. \textbf{Third Panel:} Ratio of these densities, $\rho(\phi_1)/\rho_0(\phi_1)$, together with error bars propagated forward assuming the error bars on the numerator and denominator are the square root of the number of entries in the respective bin.  Also shown in orange is the best fit top-hat function which is used to extract the gap width ($\Delta \phi_{\rm gap}$) and depth ($f_{\rm gap}$) values of this stream--subhalo encounter. Details of the error propagation and fitting procedure can be found in Appendix~\ref{sec:gap-find}. \textbf{Fourth Panel:} Sky projection of the stream as a function of $\phi_{1}$ and $\phi_{2}$. A clear underdensity can be seen around $\phi_1 \approx 50$ deg, corresponding to the location of the gap in the above panels.}
    \label{fig:example-gap}
\end{figure*}

Stream--subhalo encounters that pass both sets of cuts described above are then simulated using the \Gala{} package, details of which can be found in~\citep{GALA}. Essentially, a far more realistic version of the stream is recreated by using the~\citep{2024arXiv240801496C} particle spray method wherein the globular cluster progenitor releases test particles while orbiting the host potential. Initially, the stream evolves from a set of initial conditions chosen to roughly reproduce the position of GD-1 on the sky at redshift zero. At $t_\mathrm{pre}$, the subhalo is initialized into the simulation with its proper phase space coordinates and structural parameters. While \SatGen{} uses a truncated Green profile for the subhalo density profile, \Gala{} uses a Plummer sphere. The mass of that Plummer sphere is chosen to be equal to the total mass of the \SatGen{} subhalo at the start of the encounter, and its scale radius is chosen to be equal to the truncation radius of the \SatGen{} subhalo at that time. \apjbf{Except at small radii, the plummer sphere provides a good approximation to the enclosed mass profile of the truncated Green profile}.
\apjbf{For almost all stream--subhalo encounters,  the distance of closest approach is large enough that this is an excellent approximation}.
The stream--subhalo system is then evolved in time from $t_{\rm pre}$ to $t_{\rm post}$ within the gravitational potentials of both the host (taken from the given realization) and the subhalo.

Beyond $t_{\rm post}$, the effect of the subhalo on the stream is negligible, and it is removed from the simulation while the stream continues to evolve forward until $t_0$. As a point of comparison, for each encounter, we also create a simulated stream without the subhalo perturber. The result is an unperturbed stream that is used as a baseline for analyzing each event.

The output of each simulation is the 6D phase-space information for each of the member particles of the stream in galactocentric coordinates. We convert each of these positions and velocities to International Celestial Reference System (ICRS)  coordinates assuming that the sun is at the galactocentric Cartesian coordinates: $\vec{r}_\odot = (8.122, 0, 0.021)$~kpc with a velocity $\vec{v}_\odot = (12.9, 245.6, 7.78)$~km/s. We then transform to the stream coordinates, $\phi_1$ and $\phi_2$, defined such that the progenitor is at the origin and its velocity is entirely along the $\phi_1$ direction. To make the effect of the subhalo passage on the stream quantitative, we compute the density of the stream, $\rho(\phi_1)$, binning the number of test particles in $0.1 \deg$ intervals along $\phi_1$.

To isolate the effect of the subhalo on the density structure of the stream, we take the ratio of the density of the perturbed stream to the density of the unperturbed stream along $\phi_1$. Taking the ratio divides out density features in the stream that are inherent to the phase-space distribution of the stream progenitor and density features caused by the background gravitational potential, so that observed features are due to the effects of the subhalo alone. To actually find gaps and estimate their depths and widths, we use a modified version of the \texttt{BoxLeastSquares} method first developed for detecting exoplanet transits~\citep{2002A&A...391..369K}. Full details of the gap finding procedure are given in Appendix~\ref{sec:gap-find}. \apjbf{To summarize, we fit an inverted top-hat function to the ratio of perturbed to unperturbed densities $\rho/\rho_0$ using a $\chi^2$ analysis. This inverted top-hat function has two free parameters, $\phi_{\mathrm{gap}}$ and $\Delta \phi_{\rm gap}$, which are the central location and width of the underdense region, respectively. The depth of the best-fit top hat function is denoted $f_{\rm gap}$ and is estimated from the value of the density ratio in the gap region}. An example is shown in Figure~\ref{fig:example-gap}. The first panel shows the unperturbed density of a stream, the second panel shows the perturbed density of the same stream, and the third panel shows the ratio of these densities together with the fit of an inverted top-hat function. In the fourth panel, we show a sky projection of the perturbed stream, where the underdensity manifests as a clearly visible gap.

\section{Results}
\label{sec:results}

This section provides the main results of the study, focusing on statistical properties of gaps and the stream--subhalo encounters that cause them, as well as the frequency of gaps expected for a GD-1-like stream in MW mass host halos.

\begin{figure*}[t]
    \centering
    \includegraphics[width=2\columnwidth]{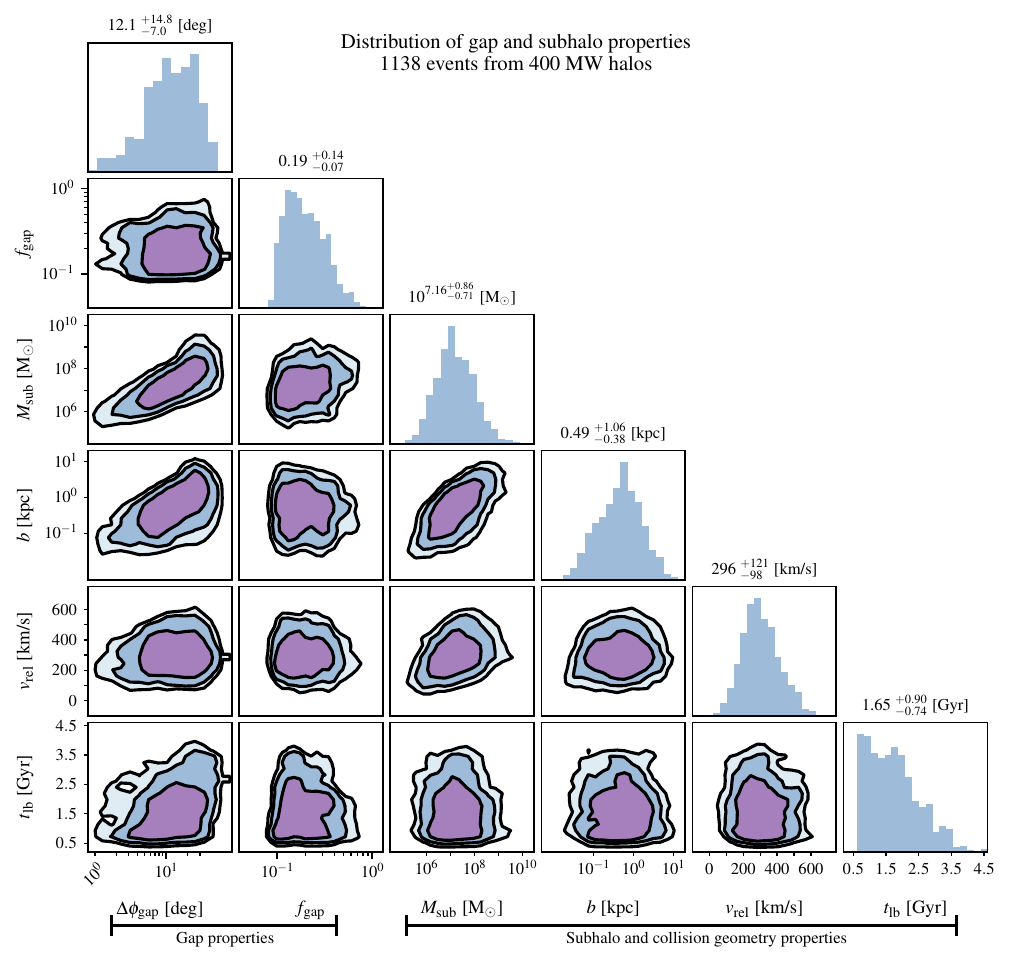}
    \caption{Joint distributions of gap, subhalo, and collision geometry properties for subhalo--stream encounters in our fiducial run (with a MW halo mass of $10^{12}~M_{\odot}$) that created gaps and passed the quality and SNR cuts defined in Appendix~\ref{sec:gap-find}. The plot variables are: gap width ($\Delta\phi$), gap depth ($f_{\rm gap}$), instantaneous subhalo mass at the time of closest approach to the stream ($M_{\rm sub}$), impact parameter ($b$), relative velocity ($v_{\rm rel}$), and time of collision ($t_{\rm lb}$). The three contours show the regions in which 63\%, 86\%, and 95\% of events are contained, respectively. The distributions contain 1138 subhalo--stream encounters from 400 realizations of a MW mass host halo.  
    }
    \label{fig:gap-corner}
\end{figure*}

\begin{figure*}[t]
    \centering
    \includegraphics[]{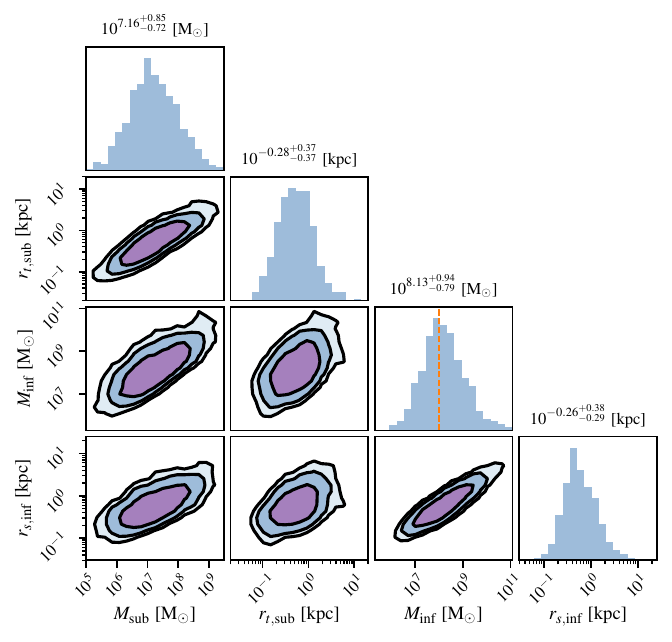}
    \caption{Joint distributions of the instantaneous subhalo mass (i.e., the subhalo mass at the time of closest approach to the stream, $M_{\rm sub}$), the tidal truncation radius ($r_{t, \rm sub}$), the subhalo infall mass ($M_{\rm inf}$), and the scale radius of the subhalo at infall ($r_{s, \rm inf}$) for the same events as Figure~\ref{fig:gap-corner}. The three contours show the regions in which 63\%, 86\%, and 95\% of events are contained, respectively. The orange dashed line at $M_{\rm inf} \approx 10^8 M_{\odot}$ corresponds to the approximate subhalo infall mass below which the subhalos are expected to be dark based on the current census of MW satellites.}
    \label{fig:subhalo-corner}
\end{figure*}

\subsection{Gap and Stream--Subhalo Encounter Properties}
\label{sec:properties_discussion}

Figure~\ref{fig:gap-corner} summarizes the gap properties and the subhalo and collision geometry properties of potentially significant stream--subhalo encounters. In particular, these are encounters whose signal-to-noise ratio (SNR) is greater than 5 together with some additional mild quality cuts to filter out spurious gaps (see Appendix~\ref{sec:gap-find} for more details). 
Shown in the figure is a corner plot of the distribution of gap widths, $\Delta \phi_{\rm gap}$, and depths, $f_{\rm gap}$. Also shown is the distribution of instantaneous subhalo mass, $M_{\rm sub}$ (i.e., the mass of the subhalo at closest approach to the stream), as well as stream-subhalo-encounter kinematics, including the impact parameter of the subhalo, $b$, its velocity relative to the stream at the point of closest approach, $v_{\rm rel}$, and also the lookback time of closest approach, $t_{\rm lb}$. 
Note that variables other than the lookback time and relative velocity are plotted on a logarithmic scale.

The one-dimensional distributions have distinct peaks. Of particular interest is the instantaneous subhalo mass distribution that peaks around $10^7\,M_\odot$ and has support down to \apjbf{$2 \times 10^6\,M_\odot$}. This highlights the potential strength of using observed gaps in objects such as GD-1 to probe dark subhalos and luminous satellites, as will be discussed further below. 

The distinct peaks in the one-dimensional distribution of $t_{\rm lb}$ correspond to pericenter passages of the stream. The encounter rate is expected to be largest near pericenter because the stream is both longest (due to its orbital modulation) and fastest at those times. Both these effects cause the stream to encounter a larger volume of potential subhalo perturbers, and increase the  probability of significant stream--subhalo encounters. Additionally, the encounter rate exhibits secular growth as the stream grows in length, and streams are sensitive to subhalo perturbations throughout their entire age. 

A number of correlations are also visible in the corner plot. First, $M_{\rm sub}$ correlates positively with $v_{\rm rel}$ and with $b$. Secondly, gap width correlates positively with $M_{\rm sub}$, $b$, and $t_{\rm lb}$. While subtle effects that are not fully understood could be at play, these correlations can partially be explained by first principle arguments, which we now discuss.

The first set of correlations ($M_{\rm sub}$ with $v_{\rm rel}$ and $b$) can be understood by considering that significant gaps can only form when $\Delta v \sim G \, M_{\rm sub}/(b \, v_{\rm rel})$ is above some threshold. If the impulse is approximately constant for most events that pass our cuts, this forces a correlation of the form $M_{\rm sub} \propto b v_{\rm rel}$ which is approximately consistent with the corner plot.

The second set of correlations (gap width with $M_{\rm sub}$, $b$, and $t_{\rm lb}$) is more involved, but can be partially understood by considering the simplified picture of gap growth presented in~\citep{Erkal:2015a} (which considered an infinitely narrow stream on a circular orbit at radius $r_0$, impacted by a Plummer sphere with scale radius $r_s$). In that study, it was shown that the growth of a gap over time can be separated into three regimes. The first regime, governed by the orbital timescale, corresponds to the initial formation of the gap within the stream. In the next phase, the gap grows over time because of particles drifting away from the impacted region, eventually forming caustics (beyond the caustic timescale defined in~\citealt{Erkal:2015a}), which sets the final phase of stream evolution. We find that the majority of stream--subhalo encounters in our sample have lookback times that are smaller than the caustic timescale, and a minority of events have larger timescales. For those with smaller lookback times, the gap width should parametrically scale as $\Delta\phi_{\rm gap} \propto (v_{\mathrm{rel}}/v_{\mathrm{rel},\perp}) \sqrt{(b^2 + r^2_s)/r^2_0}$, where $v_{\mathrm{rel},\perp}$ is the projection of the relative velocity onto the direction perpendicular to the stream. The subhalo scale radii, $r_s$, scale positively with $M_{\rm sub}$, and typical impact parameters, $b$, are also of order $r_s$. Thus, $w$ should scale positively with both $b$ and $M_{\rm sub}$. Additionally, since gaps grow with time beyond the first phase, a positive correlation is expected between $w$ and $t_{\rm lb}$, although we are not able to explain the precise scaling using the results of~\citep{Erkal:2015a}.

Appendix~\ref{sec:app-highmass} shows a similar corner plot for the high-mass realizations. Results for distributions of gap and stream--subhalo-encounter properties are similar to the fiducial run (although the peak of the subhalo instantaneous mass distribution is at a slightly larger value). This similarity in distributions highlights the insensitivity of gap properties to small changes in host virial mass discussed above and predicted by~\citep{2017MNRAS.466..628B}.

\begin{figure*}[t]
    \centering
    \includegraphics[width=2\columnwidth]{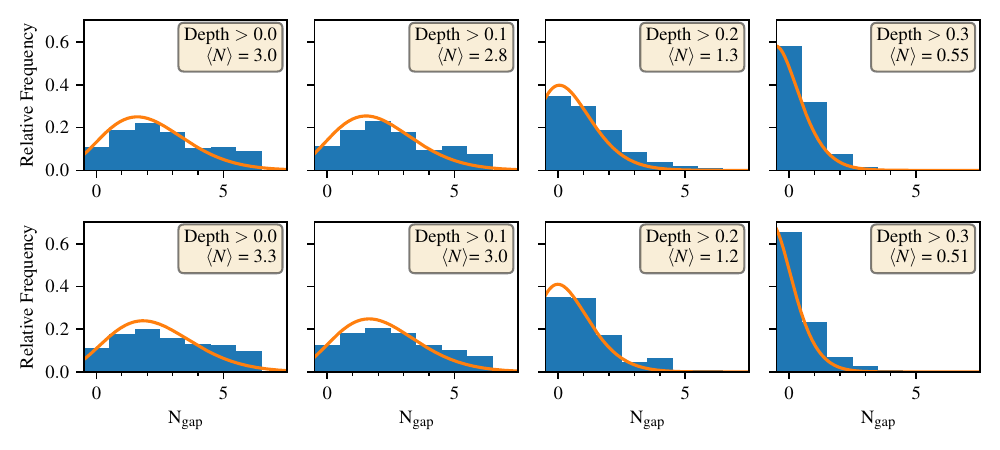}
    \caption{
    Distributions of the number of gaps that are formed in each Milky Way-like system for various thresholds of gap depth ($f_{\rm gap}$). The results approximately follow Poisson distributions with the best fits shown in orange and Poisson means denoted by $\langle N \rangle$. \textbf{Top row:} Results for the fiducial sample (with a MW halo mass of $10^{12}~M_{\odot}$). \textbf{Bottom row:} Results for the high-mass sample (with a MW halo mass of $2\times 10^{12}~M_{\odot}$).}
    \label{fig:gap-freq}
\end{figure*}

Figure~\ref{fig:subhalo-corner} shows the joint distributions of instantaneous subhalo mass, $M_{\rm sub}$, and truncation radius $r_{t,{\rm sub}}$, as well as infall mass, $M_{\rm inf}$, and infall scale radius $r_{s,{\rm inf}}$. These quantities are all positively correlated. The distribution of infall masses, which is in the $1\sigma$ range \apjbf{$\sim 2 \times 10^7 -  10^9\,M_\odot$}, is of particular interest, since it provides a direct connection to the (nonlinear) matter power spectrum and also determines the likelihood that the subhalos are luminous (dwarf satellite galaxies) or dark.
The former point is important for testing DM models that predict the power spectra different from that in the CDM model. The latter point highlights that stream gaps are potential probes of both luminous and dark  substructures,  since subhalos with peak masses (which are similar to infall masses) below $\sim 10^8 \, M_\odot$ are likely to be inefficient at forming stars. 
This is also consistent with constraints from the census of MW satellites on the infall masses of halos that host the MW satellites~\citep{Jethwa:2016gra,DES:2019ltu}, although  this may change if the Vera Rubin Observatory discovers more satellite galaxies~\citep{Nadler:2024ims}.

\subsection{Gap Frequency}

The top row of Figure~\ref{fig:gap-freq} shows histograms of the number of gaps found in each realization for our fiducial run that are above a specified depth threshold. This quantifies the expectation for how likely gap formation is in the stream, and is a key result of this work. We find that gap occurrence distributions are well fit by Poisson distributions, and show the best fit as orange curves in the figure. 
For a depth threshold of $0.1$, we find that the mean number of gaps for a GD-1-like stream is $\langle N \rangle \approx 1.8$ (the average of the best-fit Poisson distribution) for the fiducial halos. Gaps with depths above approximately $f_{\rm gap} \gtrsim 0.2$ are comparatively rare.

The bottom row of Figure~\ref{fig:gap-freq} shows the same histograms for the high mass sample. \apjbf{We find the number of gaps per realization above each depth threshold to be broadly consistent between the fiducial and high mass runs. Although the higher mass hosts have a larger number density of subhalos, the larger flyby velocities for the stream subhalo encounters lead to weaker perturbations to the stream on average, thus the high mass runs only see a modest increase in gap formation rates of about 10\%.}

Taking the observed gap near the spur in GD-1 \citep{Bonaca:2019} to be roughly 9~$\deg$ with a depth of $\sim 0.25$, it is interesting to make a rough estimate of the fraction of events that produce similar gaps in the fiducial sample. Choosing the events that produced a final gap between (7--9)~$\deg$ and a depth between $0.15$--$0.35$, we find 127 gaps in the 400 realizations, corresponding to 0.3~events per realization producing a gap similar to the gap in GD-1. Of course, \apjbf{these criteria do not select} for the presence of a spur, model the stellar population of the stream, nor quantify how observable these gaps would be in real data.  Neglecting these complications, if we focus on the existence of the gap in GD-1, our study predicts that such gaps are uncommon but not in a statistically significant way. The presence of the spur is likely to further narrow the parameter space of possible perturbers.

\section{Discussion}
\label{sec:discussion}

This section compares the results of the current study to previous literature.  It also discusses a number of limitations of the technique, some of which will be addressed in future work.  Finally, it discusses future applications of the techniques developed in this study. 

\vspace{2mm}

\subsection{Comparison to Previous Work}

Many previous studies have attempted to quantify the expectation of gap formation 
\citep[e.g.,][]{2008ApJ...681...40S, Ibata:2002, Johnston:2002, Yoon:2011,2009ApJ...705L.223C,2018JCAP...07..061B,Bonaca:2019,2024arXiv240611989M, 2025ApJ...978...79B}. Here we focus on comparing our study to results from an analysis of the FIRE simulation suite~\citep{2023MNRAS.523..428B}. We also briefly comment on results of an analytic study of gap formation done by~\citep{Erkal:2016}.

The work of \citep{2023MNRAS.523..428B} presented a comprehensive study of the dark subhalo population within FIRE, together with an analysis of expectations for subhalo encounters with a GD-1-like stream. In that study, GD-1 was modeled as a $15$ kpc cylinder, and encounter rates were found for subhalos above masses $10^6\,M_\odot$, $10^7\,M_\odot$, and $10^8\,M_\odot$ (with varying impact parameter thresholds for each group of masses). Table~\ref{tab:flyby-rates} shows a comparison between encounter rates found in~\citep{2023MNRAS.523..428B} and those found in this work using the same selection criteria in terms of subhalo mass thresholds and impact parameters. We find similar stream--subhalo encounter rates \footnote{\apjbf{It is worth noting that although \SatGen{} predicts a larger average subhalo number density than the FIRE simulations (a discrepancy previously reported in~\citep{Jiang:2020rdj}), the encounter rates in~\citep{2023MNRAS.523..428B} were calculated with an enhancement factor due to the increased subhalo population from the LMC.}}. One result of our work, that was not evaluated with the FIRE subhalo catalog, is the rate of significant gap formation following an encounter. This is shown in the last column of Table~\ref{tab:flyby-rates}. \apjbf{Since not every close flyby necessarily perturbs the stream enough to form a prominent gap, the rate of gap formation is much smaller than the flyby rate.}

\begin{table*}
    \centering
    \begin{tabular}{c c c c c}
        \toprule
         Subhalo & Impact param. & FIRE & This Work & This Work \\
         mass range & cut & Encounter rate & Encounter rate & Gap rate \\
        \midrule
         $>10^6$ $M_\odot$ & $<0.58$ kpc & 4-5 Gyr$^{-1}$ & \apjbf{$5.1$} Gyr$^{-1}$ & \apjbf{$0.38$ Gyr$^{-1}$} \\
         $>10^7$ $M_\odot$ & $<1.6$ kpc & 1-2 Gyr$^{-1}$ & \apjbf{$1.7$} Gyr$^{-1}$ & \apjbf{$0.33$ Gyr$^{-1}$} \\
         $>10^8$ $M_\odot$ & $<4.5$ kpc & 0-1 Gyr$^{-1}$ & \apjbf{$0.6$} Gyr$^{-1}$ & \apjbf{$0.1$ Gyr$^{-1}$} \\
        \midrule
        Total Rate & --- & --- & --- & \apjbf{0.81 Gyr$^{-1}$} \\
        \bottomrule
        
    \end{tabular}
    \caption{Comparison of the flyby rates between this work and those reported in \citep{2023MNRAS.523..428B}. For each threshold in subhalo mass, we consider the rate at which subhalos come within the specified impact parameter. We then determine which of these close flybys lead to the formation of a gap of depth larger than 0.1 and estimate a gap formation rate for each of these thresholds. Notably, the gap formation rate is much lower than the flyby rate. The bottom row shows the total gap formation rate for the fiducial set of realizations. Note that it is larger than the other rates, since collisions with impact parameters larger than the ones used as cuts can still form gaps.}
    \label{tab:flyby-rates}
\end{table*}
    
A number of analytic studies of stream perturbations have also been performed. In \citep{Erkal:2016}, the authors combined a numerical model of stream density evolution based on orbital perturbation theory with analytic subhalo number densities and velocity distributions. The authors modeled a GD-1-like stream on a circular orbit of radius 19~kpc, and found an expected number of gaps of 0.6 for a depth threshold of 0.25, and 0.3 for a depth threshold of 0.5. These results are similar to ours and are shown in Figure~\ref{fig:gap-freq}. However, it is worth noting that the depth thresholds used by \citep{Erkal:2016} and by us are not directly comparable, since their modeling of the stream evolution does not account for ``filling in'', the phenomena where star particles released from the progenitor after the collision can occupy the gap region, making the underdensity less extreme.

\subsection{Advantages and Limitations of this study}

The approach taken in this study has many notable advantages. One key feature is our ability to sample over a large number of realizations and easily quantify and account for effects such as halo-to-halo variance. \SatGen{} is particularly useful for this because the package enables the user to easily vary the density profiles of both DM and baryons or the mass-concentration relation. Such variations can be used to study systematics within our results; our high-mass run in this work is an example. The modular nature of our technique also allows the user to easily change any aspect of the analysis. For example, it would be straightforward to rerun our entire pipeline while varying the MW potential in which GD-1 orbits, or to model a different stream.

Our technique also has a number of limitations. One important drawback is related to differences between our stream modeling and real data. For example, 
to smoothly sample the modeled stream phase space, we run \Gala{} releasing $\approx 5\times 10^5$ test particles, which is about two orders of magnitude more than the number of real stars observed in GD-1. In addition, our technique for gap detection relies on the use of a baseline unperturbed stream model for each encounter. Dividing the binned density of perturbed streams by their counterpart unperturbed models removes unwanted features that have nothing to do with stream--subhalo encounters. However, this technique also loses contact with real data and makes the gap definition dependent on the simulations. We note that this is the standard practice in the literature and that there isn't a demonstrably better technique to use. There is also no agreed upon definition of gaps in streams. 

While the algorithm we presented in Appendix~\ref{sec:gap-find} serves as a provisional definition of gaps for this study, it is overly simplistic for the reasons discussed above. 
An important caveat of our study is that we treat each 
stream--subhalo encounter independently.
Namely, streams within our modeling do not encounter multiple subhalos and effects of one encounter do not affect any subsequent encounter. Thus, changes in stream trajectories due to subhalo encounters are not taken into account. This is a reasonable approximation, since impulsive encounters (which are those responsible for gap formation) do not significantly deflect the trajectory of a stream. Additionally, our technique does not account for cases where the stream has multiple gaps from multiple stream--subhalo encounters. These simplifications can be relaxed in future studies within the framework we have set up in this study.  \\ \\

\subsection{Further Applications of this Framework}

\textbf{Spurs and other features.} A similar analysis to that done in this study can be performed for the case of other features beyond the gap. 
For example, events that can potentially produce spur-like features could be initially evaluated by considering the impulse curve perpendicular to the stream (as opposed to the currently evaluated $\delta v_{||}$). Events with curves that predict both sizable spurs and gaps can then be simulated and the probability of forming both types of features can then be evaluated.

\textbf{Stream population studies.} This work considers stream--subhalo encounters for the case of a single, GD-1-like stream model. Instead, one could perform a similar analysis for a population of streams at various locations in phase space within a MW mass system. Such a study could indicate the types of streams most susceptible to gap (or spur) formation. Of particular interest would be a quantitative evaluation of the number of features expected for the entire stream population of the MW. This could inform us about the types of streams that are most or least susceptible to gap or spur formation, and those that can potentially be used to probe substructure below the star formation threshold and to probe deviations from the $\Lambda$CDM paradigm.

\textbf{Dark sector physics.} A natural extension of this work is to use the techniques developed in this study to probe the underlying particle theory of DM and the dark sector in which it resides. For example, the suppression of the matter power spectrum at small scales in models of warm or fuzzy DM can reduce the number count of subhalos in the mass range to which streams are sensitive. Another example involves effects of subhalo density profiles on gap (and spur) properties. Specifically, if subhalos are either more or less centrally dense than CDM predictions, this could have significant and potentially observable effects on gap width and depth, and on gap frequency.

Comparing such predictions to observations could probe various dark sector models. One classic example is models of self-interacting DM, which predict significant evolution of subhalo density profiles throughout cosmic time. Dominated by efficient heat flow, these subhalos initially form large low-density cores at their centers, which later contract into extremely high density central regions driven by gravothermal core-collapse~\citep{Balberg:2001qg, Balberg:2002ue, Koda:2011yb, Nishikawa:2019lsc, Zeng:2021ldo, Outmezguine:2022bhq, Slone:2021nqd, Jiang:2022aqw,Palubski:2024ibb}. Including dissipation in such a dark sector (with self interactions) has been shown to increase the rate of core-collapse even further~\citep{Essig:2018pzq}. Both stages of this evolution are substantially different from the CDM expectation and should result in significant variations to gap properties and production rates. \apjbf{The authors of \citep{2025ApJ...978L..23Z} have demonstrated that a core-collapsed SIDM subhalo is a reasonable candidate for the GD-1 perturber.} Building such effects into our technique will be done in future work.

\section{Conclusions}
\label{sec:conclusions}

We have developed a robust framework for modeling the impact of mostly dark subhalos on stellar streams based on a combination of semi-analytic methods to generate subhalo populations 
and dedicated stream--subhalo simulations. Our framework relaxes some assumptions used in previous studies, and was also developed in a modular fashion, with the merger tree, the subhalo orbital evolution, and the modeling of the stream each being handled separately.  This facilitates marginalization over model parameters and assumptions.  
Our framework will also allow studies of gap formation and evolution in models of dark matter other than cold dark matter.

Our main results are the following: 
\begin{itemize}
\item We find that an $\mathcal{O}(1)$ number of sizable gaps are expected to form in a GD-1-like stream over its lifetime, approximately $\sim 0.2$ of which roughly resemble the actual gap observed in GD-1. However, we do not address here how easy it is to identify such gaps in real data, and instead our gap identification relies on having access to the unperturbed stream, which allows us only to very roughly quantify the size and depth of an observed gap. Additionally, we do not model other morphological features observed in real streams, such as spurs and kinks.

\item We find that subhalos with masses $M_{\rm sub}\sim 5 \times 10^6M_\odot - 10^8 M_\odot$ at the time of closest approach with a stream are the likeliest to form significant gaps. This corresponds to subhalo masses at the time of infall of $M_{\rm inf} \sim 4 \times 10^7 M_{\odot} - 8 \times 10^8 M_{\odot}$. 
\apjbf{The current MW satellite census is consistent with all subhalos hosting galaxies down to a peak mass of $3 \times 10^8 M_{\odot}$~\citep{DES:2019ltu}. However, various galaxy formation models predict inefficient galaxy formation at lower masses, such that most halos with peak masses below $\sim 10^7 M_{\odot}$ remain ``dark''~\citep{2022MNRAS.516.3944M, 2024MNRAS.529.3387A}.} Thus, gap formation in streams can be caused by both luminous and dark substructures.

\item \apjbf{The rate of gap formation in MW-like hosts has a weak dependence on the virial mass of the host halo.}

\item The actual properties of the gaps that are formed, namely their depth and width distributions, are only mildly affected by the increased MW mass. This is consistent with the findings of~\citep{2017MNRAS.466..628B}.
\item We find a positive correlation between gap width and subhalo mass (both the instantaneous mass when the subhalo encounters the stream, and the subhalo's mass at infall), impact parameter, and the lookback time, but weak to no correlations between gap depth and these values. This could indicate that gap depth has degeneracies with the subhalo and collision-geometry properties of the flyby that created the gap.
\item We find non-trivial structure in the lookback time. The pericentric passages of the stream correspond to the largest instantaneous stream--subhalo flyby rate. We additionally observe a secular growth in the encounter rate commensurate with the growing length of the stream as it orbits in the host halo. We also find that most subhalos whose perturbations cause gaps survive until redshift zero, with only a handful being disrupted.
\end{itemize}

The modular approach developed in this work to quantify stream--subhalo encounters can be adapted and used to quantify expectations of the effect of a subhalo population on multiple stellar streams, to study stream-density power spectra observables, and to understand the statistics of gap properties in other DM models.
In the next decade, a large amount of stream observations are expected from the 
Vera Rubin Observatory, the Nancy Grace Roman Space Telescope, and the Via survey. 
The approach presented here is a necessary step in developing the tools and methodology to maximize the science gain from these observations.

\section*{Acknowledgments}

The authors would like to thank Ana Bonaca, Jo Bovy, Alyson Brooks, Fangzhou Jiang, Denis Erkal, Dylan Folsom, Kathryn Johnston, Sophia Lilleengen, Alex Riley, Sandip Roy, Robyn Sanderson, Nora Shipp, Kiyan Tavangar, Daneng Yang for useful discussions. The authors would like to thank Stony Brook Research Computing and Cyberinfrastructure and the Institute for Advanced Computational Science at Stony Brook University for access to the high-performance SeaWulf computing system, which was made possible by \$1.85M in grants from the National Science Foundation (awards 1531492 and 2215987) and matching funds from the the Empire State Development’s Division of Science, Technology and Innovation (NYSTAR) program (contract C210148).
RE acknowledges support from DOE Grant DE-SC0025309, Simons Investigator in Physics Awards~623940 and MPS-SIP-00010469, Heising-Simons Foundation Grant No.~79921, and Binational Science Foundation Grant No.~2020220.  DA is supported by DOE Grant DE-SC0009854 and Simons Investigator in Physics Awards~623940 and MPS-SIP-00010469. AP acknowledges support from Simons Investigator in Physics Awards~623940 and MPS-SIP-00010469, and the US National Science Foundation Grant PHY2210533. OS is supported by the NSF (grant No.~PHY-2210498) and acknowledges support from the Yang Institute for Theoretical Physics. OS also acknowledges support from the Simons Foundation. MK acknowledges support from NSF award 2210283.
This research was supported in part by grant NSF PHY-2309135 to the Kavli Institute for Theoretical Physics (KITP).

\appendix

\section{Rotation Curves} \label{sec:rot-curve}

\begin{figure}[h]
    \centering
    \includegraphics[width=\linewidth]{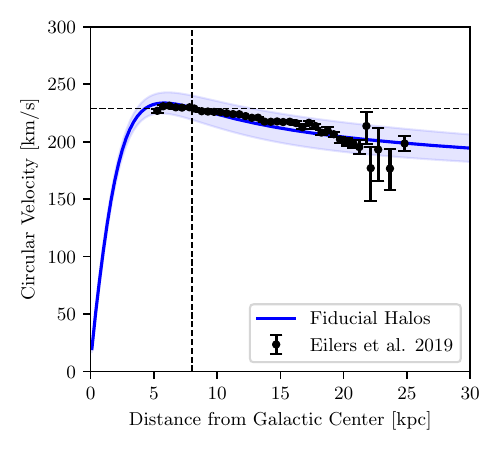}
    \includegraphics[width=\linewidth]{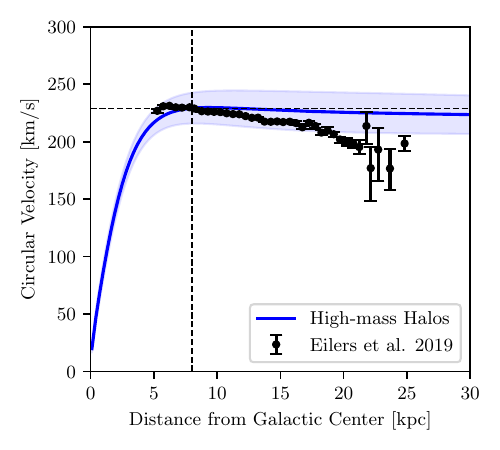}
    \caption{Rotation curves for the ensembles of halos defined in the main text. The solid line denotes the rotation curve for the mean halo properties and the blue shading around the solid line represents the halo-to-halo variance of the sample. The vertical dashed line is at 8 kpc and the horizontal dashed line at 229 km/s, corresponding to the measured distance and velocity of the Sun around the Galactic Center. \textbf{Top: } Results for the fiducial sample (with a MW halo mass of $10^{12}~M_{\odot}$), showing a good match to the measurements of~\citep{2019ApJ...871..120E}.  \textbf{Bottom: } Results for the high-mass sample (with a MW halo mass of $2 \times 10^{12}~M_{\odot}$). Unlike the fiducial run, the shape of the rotation curve deviates from the results of~\citep{2019ApJ...871..120E}, especially at larger distances.
}
    \label{fig:fiducial-rot}
\end{figure}

We have selected a value of disk mass that produces a circular velocity of $\sim 229$~km/s at a galactocentric radius of 8~kpc for the mean halo properties of our \SatGen{} ensemble. The top panel of Figure~\ref{fig:fiducial-rot} shows the rotation curves of the fiducial ensemble. We are able to get a good match to the results of \citep{2019ApJ...871..120E}. The bottom panel of Figure~\ref{fig:fiducial-rot} shows the results of the high-mass run, with parameter choices made such that we match the circular velocity value at 8~kpc. We find a worse match overall, which is expected since the heavier halo makes it impossible to match the shape of the curve past 10~kpc with just a simple disc potential. 

\section{High host halo mass results} \label{sec:app-highmass}

\begin{figure*}[h]
    \centering
    \includegraphics[width=2\columnwidth]{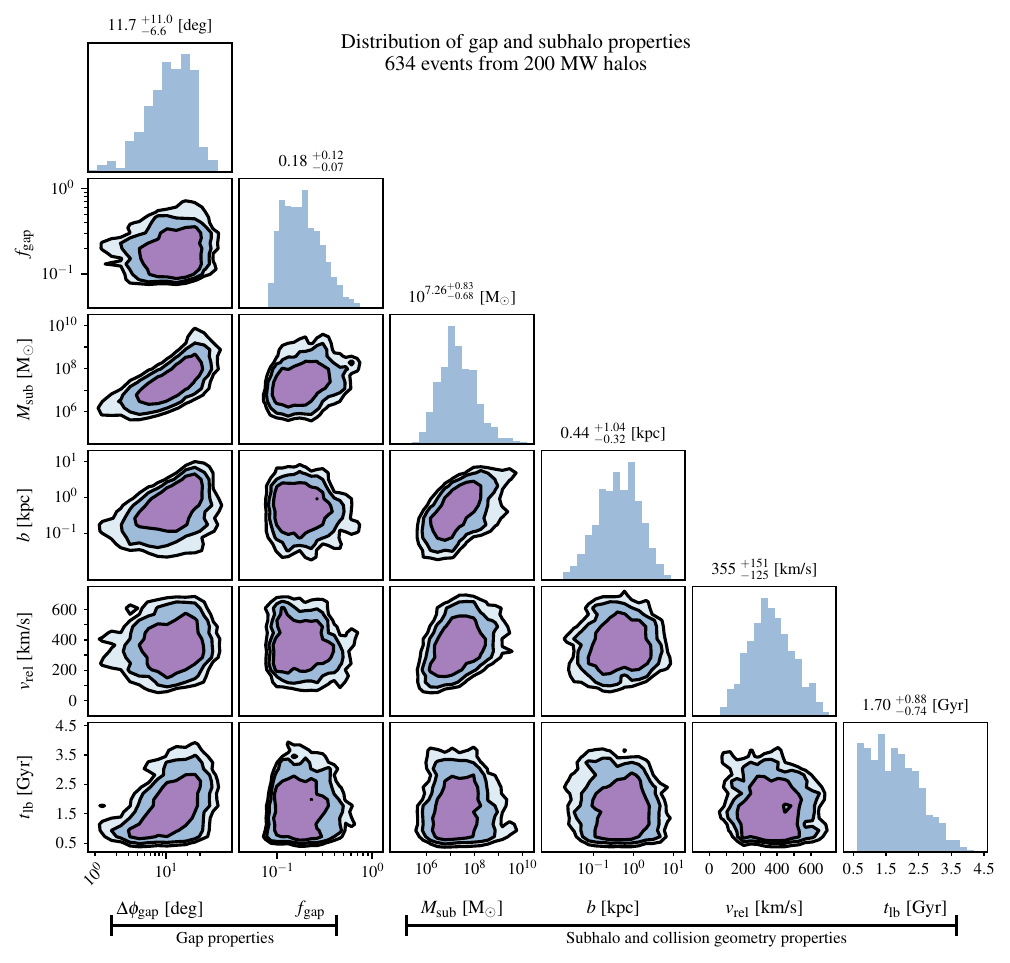}
    \caption{Joint distributions of gap, subhalo, and collision geometry properties for encounters in the high-mass run (with a MW halo mass of $2\times 10^{12}~M_{\odot}$) that created gaps and passed the quality and SNR cuts defined in Appendix~\ref{sec:gap-find}.  See Figure~\ref{fig:gap-corner} for the definitions of all variables. We find similar sets of correlations as in the fiducial run, with very similar distributions of gap properties, consistent with \citep{2017MNRAS.466..628B}. We find that the mass distribution peaks at a slightly higher value than the fiducial run, and that the relative velocity distribution is wider and peaked at a higher value, as expected from the increased virial velocity due to a heavier host mass.
    }
    \label{fig:gap-corner-highmass}
\end{figure*}

Figure~\ref{fig:gap-corner-highmass} shows the joint distributions for our high-mass sample. The correlations are qualitatively similar to those observed in the fiducial sample, which are discussed in further detail in Section~\ref{sec:properties_discussion}. 

\section{Gap Finding} \label{sec:gap-find}

To automatically find gaps in the output of the N-body simulations, we use an algorithm based on the \texttt{BoxLeastSquares} method from \citep{2002A&A...391..369K}. First, we define the binned linear densities, $\rho_{i,0}(\phi_1)$ and $\rho_i(\phi_1)$, for the unperturbed and perturbed streams, respectively, as the number count of test particles within the i'th $\phi_1$ bin. Assuming each of these bins is independent, we take $\delta \rho_{i,0} = \sqrt{\rho_{i,0}}$ and $\delta \rho_i = \sqrt{\rho_i}$ as errors. We then restrict attention to bins in which both $\rho_i$ and $\rho_{i,0}$ are greater than 10, so that we can model the errors as Gaussian, and compute the ratio $f_i = \displaystyle\frac{\rho_i}{\rho_{i,0}}$ with error $\delta f^2_i = \displaystyle f^2_i\left[\left(\delta \rho_i/\rho_i\right)^2 + \left(\delta \rho_{i,0}/\rho_{i,0}\right)^2 \right]$.

Next, we turn to the gap finding procedure. We first define the null hypothesis by evaluating the weighted average of the bins,
\begin{equation}
    \bar{f} = \frac{\sum_i f_i/\delta f^2_i }{\sum_i 1/\delta f^2_i}\ ,
\end{equation}
and computing,
\begin{equation}
    \chi_{\mathrm{null}}^2 = \sum_i \frac{(\bar{f} - f_i)^2}{\delta f^2_i}.
\end{equation}

We model the gap as a family of piecewise functions with two free parameters: the central location of the gap $\phi_\mathrm{gap}$ and the width of the gap in degrees $\Delta\phi_{\rm gap}$. The functional form is given by
\begin{equation}
f(\phi_1) =  
\begin{cases} 
      f_{\mathrm{out}} & |\phi_{\mathrm{gap}} - \phi_1| > \Delta\phi_{\rm gap}/2 \\
      f_{\mathrm{in}} & |\phi_{\mathrm{gap}} - \phi_1| < \Delta\phi_{\rm gap}/2. \\
\end{cases}
\end{equation}
The parameters $f_{\mathrm{out}}$ and $f_{\mathrm{in}}$ are estimated from the weighted means of the $f_i$ outside the gap region and inside the gap region, respectively. We then scan over a grid where $\phi_{\mathrm{gap}}$ can be centered on any of the midpoints between the bins, and the width can be between 0.2~deg and 30~deg, in steps of 0.1~deg. For each of these putative gaps, we compute
\begin{equation}
    \chi_{\mathrm{gap}}^2 = \sum_{i \in \mathrm{gap}} \frac{(f_{\mathrm{in}} - f_i)^2}{\delta f^2_i} + \sum_{i \notin \mathrm{gap} } \frac{(f_{\mathrm{out}} - f_i)^2}{\delta f^2_i}, 
\end{equation}
and record the model parameters of the gap model with the smallest $\chi_{\mathrm{gap}}^2$. We then define the depth of the gap as $f_{\rm gap} =f_{\mathrm{out}} - f_{\mathrm{in}}$ and $\delta f_{\rm gap} = \sqrt{\delta f^2_{\mathrm{out}}+ \delta f^2_{\mathrm{in}} }$. Finally, two cuts are placed on the gaps that are found. First, we require $\Delta \chi^2 = \chi_{\mathrm{null}}^2 - \chi_{\mathrm{gap}}^2$ to be larger than the number of bins included in the gap region. Second, defining a ``signal-to-noise'' ratio for the gap as $\mathrm{SNR} = \displaystyle f_{\rm gap}/\delta f_{\rm gap}$, we require $\mathrm{SNR} > 5.0$. We have confirmed that these effectively filter out spurious gaps from our sample.

\newpage

\bibliography{biblio}{}
\bibliographystyle{aasjournal}

\end{document}